\documentclass[aps,prb,showpacs,superscriptaddress,amsmath,amssymb,floatfix]{revtex4}
\usepackage{graphicx}
\usepackage{dcolumn}
\usepackage{bm}
\usepackage{stmaryrd}
\usepackage{latexsym}
\usepackage{amssymb}
\usepackage{amsfonts}
\usepackage{amsmath}
\usepackage{fancybox}
\usepackage{color}

\begin{document}
\title{Conductance of bilayer graphene in the presence of a magnetic field:
Effects of disorder}
\author{H. Hatami}
\affiliation{Department of Physics, University of Tabriz, Tabriz
51665-163, Iran}
\author{N. Abedpour}
\affiliation{School of Physics, Institute for Research in
Fundamental Sciences, (IPM) Tehran 19395-5531, Iran}
\author{A. Qaiumzadeh}
\affiliation{School of Physics, Institute for Research in
Fundamental Sciences, (IPM) Tehran 19395-5531, Iran}
\affiliation{Institute for Advanced Studies in Basic Sciences
(IASBS), Zanjan 45195-1159, Iran}
\author{Reza Asgari}
\email{asgari@ipm.ir}
\affiliation{School of Physics, Institute for Research in
Fundamental Sciences, (IPM) Tehran 19395-5531, Iran}
\begin{abstract}
We investigate the electronic transport properties of unbiased
and biased bilayer graphene nanoribbon in n-p and n-n junctions
subject to a perpendicular magnetic field. Using the non-equilibrium Green's function method and the Landauer-B\"{u}ttiker
formalism, the conductance is studied for the cases of clean, on-site, and edge disordered bilayer graphene. We show that the lowest Hall plateau remains unchanged in the presence of disorder, whereas asymmetry destroys both the plateaus and conductance quantization. In addition, we show that disorder induces an enhancement of the conductance in the n-p region in the presence of magnetic fields.
Finally, we show that the equilibration of  quantum Hall edge states between distinctively doped regions causes Hall plateaus to appear in the regime of complete mode mixing.

\end{abstract}
\pacs{73.63.-b, 72.80.Ng, 81.05.ue, 05.60.Gg}

\maketitle
\section{Introduction}

In recent years, after the success of fabrication of  both monolayer~\cite{novoselov} and multilayer graphene sheets~\cite{berger,novo,zhang}, there has been a lot of interest in the transport properties of graphene nano-ribbons, especially the
behavior of low-energy charge carrier excitations~\cite{monolayer,bilayer,bilayer0}. Ideal monolayer graphene is a gapless semimetal with zero density of states at the Dirac points. The low-energy electronic excitations in the vicinity of the Dirac points have linear dispersions, and are described by an effective massless Dirac Hamiltonian. The low-energy electrons in bilayer graphene, on the other hand, have a quadratic dispersion relation. Both for monolayer and bilayer graphene, the wave functions
 are composed of two sublattices $A$ and $B$, and give rise to the
chirality of the charge carriers. Therefore, the charge carriers in graphene are chiral and has awaken an enormous interest in graphene, for instance with regards of effects such as the Berry phase, which is $\pi$ in monolayer graphene and $2\pi$ in bilayer graphene.
Although intrinsic bilayer graphene is a zero-gap semimetal, it exhibits very interesting properties when a gate
voltage is applied, which makes bilayer graphene into a tunable band gap semiconductor~\cite{kuzmenko, mak}. The band gap determines the threshold voltage and the on-off ratio of field effect transistors and diodes. Therefore, bilayer graphene is more convenient for applications in nano-electronic industry than monolayer graphene~\cite{bilayer1}.

One of the exotic phenomena that has been observed in monolayer and bilayer graphene is the anomalous quantum Hall effect
~\cite{novoselov,bilayer2}. The nature of massless chiral Dirac charge carriers in monolayer graphene gives rise to this property of the
Hall plateaus, that behave as $\sigma_{xy}=\pm\sigma_0 (N+1/2)$, with $N$ being the Landau level index and $\sigma_0=4e^2/h$. The factor of $4$ originates from the valley and spin degeneracies. In undoped bilayer graphene, the sequence of Hall plateaus, with
$\sigma_{xy}=\pm \sigma_0 N$, were observed. The first plateau at $N=0$ is missing which implies that bilayer graphene is metallic
at the neutrality point, while the standard quantum Hall effect in bilayer graphene can be recovered by applying a gate voltage. The
quantum Hall states, fully quantized due to the presence of a magnetic field, as well as broken-symmetry states at intermediate filling factors such $0$, $\pm1$, $\pm2$ and $\pm3$, were experimentally observed by Feldman~\textit{et al.}~\cite{feldman}.

In a perfect nanoribbon, the electron transmission via subbands due to lateral confinement of the electronic states implies
the quantization of the conductance in units of $G_0=2e^2/h$.\cite{Ihnatsenka} Recently, the zero-temperature conductance of free-disordered monolayer and unbiased bilayer graphene nanoribbons in the presence of a uniform perpendicular
magnetic field was calculated~\cite{xu}. The conductance in monolayer graphene nanoribbon is given by $2(n+1/2)G_0$ for the case of
zigzag edges, and $nG_0$ for the case of armchair edges. On the other hand, it was shown that in a bilayer graphene nanoribbon the conductance is quantized as $2(n+1)G_0$ for zigzag edges, and $nG_0$ for armchair edges, where $n$ is an integer.

The quantum Hall effect and quantized transport in graphene junctions in the bipolar (p-n), and unipolar (n-n or p-p) regimes was
investigated theoretically and experimentally by several groups~\cite{long, abanin}. Long,~\textit{et al.}~\cite{long} by using  the Landauer-Buttiker formalism, showed that on-site disorder induces the enhancement of the transport in monolayer graphene p-n junctions in the presence of a magnetic field. On the other hand, they showed that in the n-n junction, the lowest plateau survives in a sufficiently broad range of on-site disorder strengths. They also showed that in a particular range of disorder strengths new plateaus (\emph{i.e.}
$G=3e^2/h$ and $e^2/h$) emerge~\cite{long,li}, something also observed experimentally.

Transport measurements in high quality bilayer graphene pnp junctions have also been performed, and electron mobilities up to $10000$ cm$^2$/(V s) have been measured for gapless systems, and an on-off ratio up to $20000$ for gapped systems.~\cite{pnp} Moreover, the
fractional-valued quantum Hall plateaus due to equilibration of quantum Hall edge states between distinctively doping regions have been observed .~\cite{pnp} Consequently, the conductance exhibits plateaus arising from the mixing of edge states at the interfaces.

In this paper, we investigate the conductance of the chiral massive carrier  in the presence of a uniform
perpendicular magnetic field, both for unbiased and biased bilayer graphene nanoribbons configured as n-n and n-p junctions~\cite{pnp,pn}. In addition, we study the influence of on-site and edge disorders on the conductance. For these purposes,
we use the tight-binding model and Landauer-B\"{u}ttiker formalism together with the non-equilibrium Green's function
approach~\cite{datta}. It should be noted that, in general, the absolute magnitude of the magnetic filed can cause reflection at the boundaries of the electronic devices~\cite{baranov}. We restrict our attention in this article to the case of a system in which the reflection on the boundaries due to the magnetic field can be ignored.

The paper is organized as follows. In Sec. II, we introduce our model and formalism, such as the tight-binding Hamiltonian for
bilayer graphene junction, and a recursive method for calculating the Green's function. In Sec. III, our numerical results for the
conductance of disordered and biased bilayer graphene junctions in the presence of magnetic field are presented. Finally, we conclude
in Sec. IV with a brief summary.

\section{Model and method}

\begin{figure}
\includegraphics[width=8.5cm]{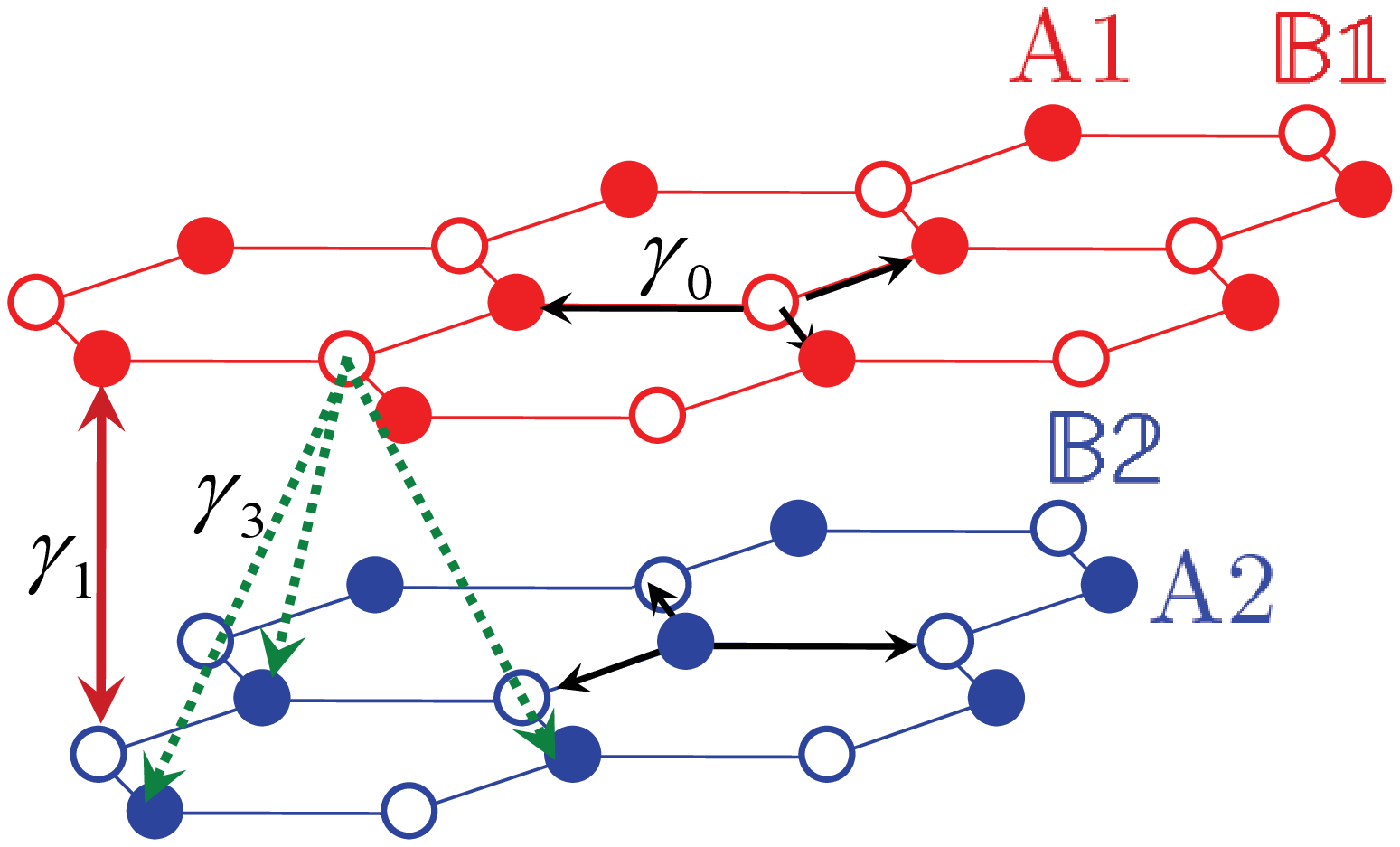}
\includegraphics[width=8.5cm]{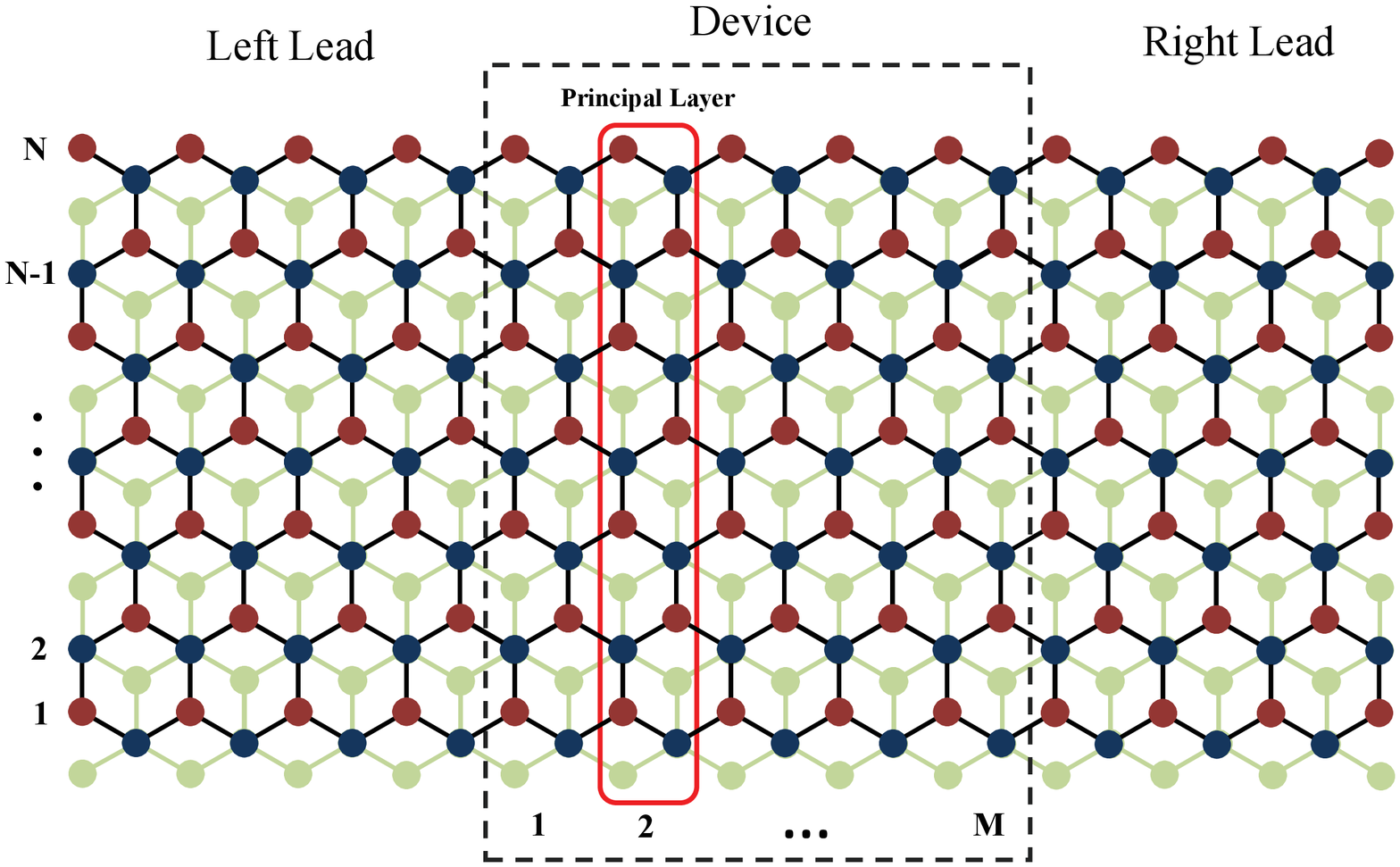}
\caption{(Color online)(Left): Schematic 3D view of bilayer graphene
incorporating all coupling energies. (Right): Schematic picture of
zBGNR with length $M$ and width $N$ atoms, and the definition of the principal
layer.}\label{bilayer}
\end{figure}

We consider a bilayer graphene ribbon with Bernal stacking
(AB)~\cite{bernal}, as a conductor connected to the left and the
right leads as depicted in Fig.~\ref{bilayer}.  The model Hamiltonian
is given by,
\begin{equation}
H=H_{center}+H_L+H_R
\end{equation}
where $H_{center}$, $H_L$ and $H_R$ are the Hamiltonian of the
center region, the left and the right leads, respectively. Two leads
are considered as perfect semi-infinite bilayer graphene
nanoribbons. Notice that the leads are also structured by the Bernal stacking bilayer graphene. We consider the nearest-neighbor tight-binding
Hamiltonian with one $\pi$-orbital per site on the lattice. The
effective one-body Hamiltonian of bilayer graphene in the presence
of the magnetic field is given as follows,
\begin{eqnarray}
H&=&-\gamma_{0}\sum_{l,\langle
i,j\rangle}(e^{i\phi_{i,j}}a^{\dagger}_{l,i} b_{l,j}+h.c.)\\
\nonumber&-&\gamma_{1}\sum_{i}(a^{\dagger}_{1,i}
b_{2,i}+h.c.)-\gamma_{3}\sum_{\langle
i,j\rangle}(e^{i\phi_{i,j}}b^{\dagger}_{1,i} a_{2,j} + h.c.)\\
\nonumber &+&\sum_{l,i}v_{l}( a^{\dagger}_{l,i} a_{l,i}+
b^{\dagger}_{l,i} b_{l,i})\\ \nonumber
&+&\sum_{l,i}\big[(w_i+(-1)^l\Delta)(a^{\dagger}_{l,i}
a_{l,i}+b^{\dagger}_{l,i} b_{l,i})\big]
\end{eqnarray}
where $a^{\dagger}_{l,i}$ and $a_{l,i}$ ($b^{\dagger}_{l,i}$ and
$b_{l,i}$) are the creation and annihilation operators at sublattice
$A$ ($B$) in the layer $l=1, 2$ at the $i$th site, respectively. The
intralayer nearest-neighbor hopping energy is $\gamma_{0} = 3.16
eV$, the hopping energy between on-top sublattices $A$ and $B$ in
different layers is $\gamma_{1} = 0.39 eV$ and furthermore,
$\gamma_{3} = 0.315 eV$ denotes the hopping energy between not
on-top sublattices $A$ and $B$ between two layers~\cite{kuzmenko}.
Another hopping energy between the nearest-neighboring layers,
$\gamma_{4} = 0.04 eV$, is very small compare to $\gamma_{0}$ and
can be ignored. In the presence of the external perpendicular
magnetic field ${\bf B}$, the hopping integral acquires the Peierls
phase factor given by $e^{i\phi_{i,j}}$ where $\phi_{i,j}=\int_{i}^{j}
{\bf{A}}\cdot d {\bf{l}}/\phi_{0}$, with the magnetic flux quantum
$\phi_{0}=\hbar/e$. We use the Landau gauge as ${\bf{A}}=(-By, 0,
0)$. The applied magnetic field is considered to be only on the center
region. $v_{l,i}$  reduces to the bias voltage $E_{L}$ ($E_{R}$) on
the left (right) lead and can be controlled by the gate voltage. The
electrostatic potential changes from the right lead to the left lead
and is assumed to be linear as $v_{l}=k(E_{R}-E_{L})/(M+1)+E_{L}$,
$k=1, 2, ..., M$, where $M$ is the length of the center region (see
Fig.~1). We consider on-site disorder $w_{i}$, being a random
variable with a uniform distribution in an interval $[-W/2,W/2]$
with the disorder strength $W$ which exists only in the center
region. The size of the central region, \textit{i.e.} conductor, is
given by $4N\times M$ atoms. Here we define the asymmetric between
two layers, $\Delta$, indicating the difference between on-site
energies. The current can be calculated from the
Landauer-B\"{u}ttiker formula~\cite{datta} as
\begin{equation}
I = \frac{2e}{h}\int d\epsilon
T_{LR}(\epsilon)[f_{L}(\epsilon)-f_{R}(\epsilon)]
\end{equation}
where
$f_{\alpha}(\epsilon)=1/({\exp[(\epsilon-eV_{\alpha})/k_{B}T]+1})$,
$(\alpha = L, R)$, is the Fermi distribution function in the leads.
To calculate the transmission coefficient $T_{LR}(\epsilon)$, we use
\begin{equation}
 T_{LR}(\epsilon)= Tr[\Gamma_{L}G\Gamma_{R}G^{\dag}]
\end{equation}
where the line width function $\Gamma_{\alpha}$, that describes the
coupling between the conductor and the leads is given by
\begin{equation}
\Gamma_{\alpha}(\epsilon)=i[\Sigma_{\alpha}^{r}(\epsilon)-\Sigma_{\alpha}^{a}(\epsilon)],
\end{equation}
in which the retarded Green's functions is written as
\begin{equation}
G(\epsilon)=\frac{1}{\epsilon-H_{center}-\Sigma_{L}^{r}(\epsilon)-\Sigma_{R}^{a}(\epsilon)}
\end{equation}
The retarded self-energy $\Sigma_{\alpha}^{r}$, due to the coupling
to $\alpha$-th lead is
\begin{eqnarray}
\Sigma_{L}^{r}=h^{\dag}_{LC}(\epsilon-H_{L})^{-1}h_{LC}\\ \nonumber
\Sigma_{R}^{r}=h_{CR}(\epsilon-H_{R})^{-1}h^{\dag}_{CR}
\end{eqnarray}
where $h_{LC}$($h_{CR}$) is the hopping Hamiltonian from the left
lead to the center region (from center region to the right lead) and
$G_{\alpha}=(\epsilon-H_{\alpha})^{-1}$ can be calculated by an
iterative method numerically~\cite{selfenergy1,selfenergy2}. We
assume an infinite stack of principal layers with the
nearest-neighbor interactions. A principle layer is defined as the smallest group of neighbouring atoms planes such way that only nearest-neighbour interactions exist between principle layers ( see Fig.~1). Thus, we can transform the original
system into a linear chain of the principal layers. By using this
approach, we write the matrix elements of $(\epsilon-H)G=1$ in
the following form~\cite{Nardelli}
\begin{eqnarray}\label{green1}
(\epsilon-H_{00})G_{0,0}&=&1+H_{01}G_{1,0}\\ \nonumber
(\epsilon-H_{00})G_{1,0}&=&H_{01}^{\dag}G_{0,0}+H_{01}G_{2,0}\\
\nonumber ...\\ \nonumber
(\epsilon-H_{00})G_{n,0}&=&H_{01}^{\dag}G_{n-1,0}+H_{01}G_{n+1,0}
\end{eqnarray}
in which $H_{00}$ and $H_{01}$ describe the coupling within the
principal layer and the adjacent principal layers, respectively. For
simplicity, we assume that $H_{00}=H_{11}=H_{22}...$ and
$H_{01}=H_{12}=H_{23}...$. Notice that $G_{nm}$ is the matrix element of
the Green's function between the principal layers. It is easy to
obtain an iterative set of equations for $G_{n,0}$
\begin{equation}\label{g}
G_{n,0}=t_{i}G_{n-2^{i},0}+\widetilde{t}_{i}G_{n+2^{i},0}
\end{equation}
for $n\geq 2^{i}$, where
\begin{eqnarray}
t_{i}=(1-t_{i-1}\widetilde{t}_{i-1}-\widetilde{t}_{i-1}t_{i-1})^{-1}t_{i-1}^{2}\\
\nonumber
\widetilde{t}_{i}=(1-t_{i-1}\widetilde{t}_{i-1}-\widetilde{t}_{i-1}t_{i-1})^{-1}\widetilde{t}_{i-1}^{2}
\end{eqnarray}
where
\begin{eqnarray}
t_{0}=(\epsilon-H_{00})^{-1}H_{01}^{\dag}\\
\nonumber
\widetilde{t}_{0}=(\epsilon-H_{00})^{-1}H_{01}
\end{eqnarray}
then we can write
\begin{eqnarray}
G_{1,0}&=&t_{0}G_{0,0}+\widetilde{t}_{0}G_{2,0}\\ \nonumber
&=&(t_{0}+\widetilde{t}_{0}t_{1})G_{00}+\widetilde{t}_{1}G_{4,0}\\
\nonumber ... \\ \nonumber
&=&(t_{0}+\widetilde{t}_{0}t_{1}+...+\widetilde{t}_{0}...\widetilde{t}_{n-1}t_{n})G_{0,0}+\widetilde{t}_{n}G_{2^{n+1},0}
\end{eqnarray}
We solve Eq.~(\ref{g}) iteratively. This process is repeated
until $t_{n+1},{\widetilde{t}}_{n+1}<\varepsilon$, in which $\varepsilon$ is a tiny value and is chosen as small as one
pleases. Therefore $G_{2^{n+1},0}\simeq 0$, and the transfer matrix
is thus given by
$T=t_{0}+\widetilde{t}_{0}t_{1}+\widetilde{t}_{0}\widetilde{t}_{1}t_{2}+...+\widetilde{t}_{0}...\widetilde{t}_{n-1}t_{n}$.
Accordingly, we can write $G_{1,0}=TG_{0,0}$ and $G_{0,0}=\bar {T}G_{1,0}$.
The self-energies of the conductor-leads are
$\Sigma_L=H_{01}^\dagger\bar T$ and $\Sigma_R=H_{01}T$. Finally, the
zero-temperature conductance $G=\lim_{V\rightarrow 0}\frac{dI}{dV}$,
can be obtained by using the Landauer-B\"{u}ttiker formalism as
\cite{datta}
\begin{equation}
G=\frac{2e^2}{h}T_{LR}(\epsilon_F).
\end{equation}

\section{Numerical results}

In this Section, we present our numerical results for zero-temperature conductance of unbiased and biased zBGNR in the
presence of a magnetic field as well as various types of disorders. We assume that the width of the nanoribbons is $N=45$,
the voltage of the left lead is $E_L=-0.2$. All the energies are in units of $\gamma_0$. We neglect the effect of the Zeeman splitting and the spin-orbit interaction, which are important only at very low energies, at which disorder effects normally dominate.\cite{monolayer} We define dimensionless magnetic field as $\phi\equiv(3\sqrt{3}/4)a^2B/\phi_0$, where $2\phi$ is the magnetic flux in a honeycomb lattice.

In connection with the formation of the Hall plateaus, it is necessary to consider ribbons with width greater than the
magnetic length scale, $l_B=\sqrt{\hbar/eB}$. We consider $\phi=0.01$ corresponds to $l_B\approx 15{\AA}$ which
is smaller than the considered ribbon size with width $L_y(N=45)\approx10$nm, and length $L_x(M=21)\approx 5.5$nm.

The conductance of unbiased clean zBGNR as a function of $E_R$ is shown in Fig.~2 for various sizes, both in the absence ($\phi=0$) and presence ($\phi=0.01$) of a magnetic field. In the absence of a magnetic field, the conductance in the n-n region ($E_R<0$) is
quantized due to transverse confinement of the ribbon, and well described by $G=2(n+1)G_0$, with the minimum conductance of a zBGNR being $2G_0$~\cite{xu}. Moreover, the conductance is independent of the ribbon length at low $E_R$ values. As one
can see in the n-n region for $\phi=0$, the energy spacings between plateaus are not equidistant (whereas in monolayer graphene they are), because of the quadratic dispersion relation. The widths of the conductance steps are related to the energy scale between the successive modes in the energy spectrum. Therefore the conductance is sensitive to $E_L$ values, and the number of plateaus increases with increasing bias voltage, \textit{i.e.} $|E_L-E_R|$. Also for $E_R<E_L$ there are no plateaus, identical to the case of monolayer graphene~\cite{long}. In the n-p region, $E_R>0$, the conductance occurs due to the chiral charge carriers tunneling between n and p
regions, and the conductance is always less than the corresponding plateau value in the n-n region. In this region the conductance decreases with increasing length of the ribbon $M$, since the number of scattering centers increases.

The effect of a high magnetic field, $\phi=0.01$ on the conductance of a clean graphene junction
in unipolar and bipolar regimes is shown in Fig.~2. In the n-p region the length dependence of the Peierls phase factor gives rise to a non-monotonic behavior of the conductance as a function of the length and energy $E_R$, noticeably at very low $E_R$ values. This behavior is in contrast with the result obtained for the zero magnetic field. Our results show that the conductance of the clean sample in bipolar
regime is suppressed dramatically in the presence of the magnetic field. However, the first and the second Hall plateaus survive (\textit{i.e.} $G/G_0=2$ and $4$) in the n-n region.

We also study the effect of asymmetry between two layers, $\Delta$, \textit{i.e.} when the two layers have different on-site energies.
The asymmetry here leads to the opening of a gap between the conduction and valance bands. In Fig.~3, we plot the conductance of a free-disordered zBGNR as a function of $E_R$ in the absence of the magnetic field. In the n-n region, asymmetry leads to a decrease of the conductance, while in the n-p region, asymmetry results in an enhancement of the conductance. This effect can be described based on the channels of the charge carriers. In the n-n region, electrons are only charge carriers while in the n-p region, because of the existence of asymmetry between two layers, one layer is n-doped and the other is p-doped.
Accordingly, both electrons and holes play a role in the transport. In the n-n region, the conductance fluctuations occur for low $|E_R|$ and
increase with length size $M$. Importantly, opening a gap affects the transversal confinement and the quantized steps are destroyed by asymmetry in the n-n region.

The conductance of a zBGNR for various $\Delta$ is plotted in Fig.~4. The asymmetry leads to an increase of the conductance in the
n-p region, and a decrease in the n-n region. The conductance fluctuations increase with increasing $\Delta$ values. In the right
panel of Fig.~4 we show the effect of the magnetic field on the conductance in the presence of asymmetry. In the n-n region the Hall
plateaus are destroyed by asymmetry and accordingly the conductance reduces. On the other hand, in the n-p region asymmetry leads to increasing conductance.

We are now in the position to introduce some disorder and study the effects of disorder on the conductance of zBGNR in the presence of
a magnetic field B. We consider the effect of on-site disorder on the conductance with a uniform disorder distribution in the range of $[-W/2,W/2]$. In Fig.~5, we plot the conductance as a function of $E_R$ for the various $W$ for $\phi=0$ (left panel). The number of realizations is $200$. In the n-n region, the conductance is suppressed. In the n-p region at low $E_R$ values, the conductance increases when the
strength of disorder increases. The conductance is independent of disorder in the large positive $E_R$ regions.

We also consider a disordered zBGNR in the presence of the high perpendicular magnetic field. The quantum Hall effect in gapless bilayer graphene occurs at the filling factors $\nu=\pm1, \pm2, \pm3 ...$, in which there are $|\nu|=|n_{0}h/eB|$ edge modes propagating in the opposite directions at $\nu>0$ and $\nu<0$. Here $n_{0}$ is the charge density. The conductance plateaus for unipolar regime is given by
$G_{nn}/G_0=G_{pp}/G_0=2(\min(|\nu_L|,|\nu_R|))=2, 4, 6 ...$ and for a bipolar system the conductance is $G_{pn}/G_0=2(\frac{|\nu_L||\nu_R|}{|\nu_L|+|\nu_R|})=1, \frac{4}{3}, \frac{3}{2}, 2 ...$ in the two-terminal ohmic regime.~\cite{abanin,pnp} It is worthwhile mentioning that those expressions can also be applied to monolayer nanoribbons in the ohmic regime since these two materials have similar resistivities and thus also similar mean free paths~\cite{aristizabal}.

For the unipolar regime the edge states which are common between the left and right regions propagate between the two leads, while $|\nu_L-\nu_R|$ states do not contribute to the conductance. In the bipolar regime, on the other hand, mode mixing occurs in the interface of two regions and for complete mode mixing the conductance plateaus obey the aforementioned formula for $G_{pn}/G_0$. In unipolar regime our numerical calculations show that ($\phi=0.01$, right panel) the lowest Hall plateaus remain unchanged in the presence of small disorder strengths, whereas the Hall plateaus are destroyed with increasing disorder strengths. We see perfect Hall plateaus in the unipolar regime with no equidistance in the scale of $E_R$. In the clean bipolar regime the Hall edge states are separated for electrons and holes, and leads to suppression of the conductance. In addition, small strengths of on-site disorder induce the enhancement of the conductance of zBGNR in the presence of a magnetic field in the n-p region. At small strengths of disorder, $W<1$, in the n-p region, the
conductance is enhanced due to the mixture of electron and hole edge states. Thus in the bipolar regime, mode mixing at interface leads to two- terminal conductances. On the other hand, for large values of disorder strength, the system enters the insulating regime and the conductance is very small for all $E_L$ and $E_R$. We expect that the lowest Hall plateau survives only within certain range of disorder strengths. In the inset of Fig.~5 the conductance is shown in the ohmic regime and it obeys $G_{pn}/G_0=2(\frac{|\nu_L||\nu_R|}{|\nu_L|+|\nu_R|})=1, \frac{4}{3}, \frac{3}{2}, 2 ...$, the edge state equilibration condition. It should be noted that the ohmic behavior has been observed experimentally in bilayer graphene pnp junctions as well as graphene p-n junctions~\cite{abanin,pnp}. Our numerical results are in excellent agreement with the recent experiment.

We also investigate the effect of asymmetry in the presence of on-site disorder and our results are shown in Fig.~6. The Hall plateaus are
destroyed by asymmetry even for very low disorder strengths. Also, asymmetry destroys finite size quantization of the conductance in
the n-n region. The strong fluctuations vanish in the n-n region for small $E_R$ values and in the presence of disorder, as is shown in Fig.~6 for the case of $\phi=0$.

Another type of disorder which is indispensable in real nanoribbons is edge disorder~\cite{ding}. This type of disorder is generated by
eliminating carbon atoms randomly along the edges of GNR. Note that because of our limitations in this approach, we consider only one
layer depth edge disorder. In Fig.~7, we have shown the effect of edge disorder on the conductance of zBGNR as a function of $E_R$ for
two different nanoribbon lengths $M$. The conductance increases in the n-p region at low $E_R$ as compared to the clean system,
and it is independent of disorder in large $E_R$. It is important to investigate the persistence of the Hall conductance plateaus versus
edge disorder. As we have shown in Fig.~7, the Hall plateaus remain unchanged in the presence of edge disorder as well as on-site
disorder for $\phi=0.01$.

\section{Summary and Conclusion remarks}

In summary, we studied the effect of on-site and edge disorder on the conductance of biased zigzag bilayer graphene nanoribbon
subject to a uniform perpendicular magnetic field. Our approach was based on the non-equilibrium Green's function method and
Landauer-B\"{u}ttiker formalism. Our results show that the lowest Hall plateaus can survive in the presence of a broad range of disorder strengths in the n-n region, while an asymmetry between two layers destroys them. On the other hand, disorder induces an enhancement of the conductance in the presence of the magnetic field in the n-p region.
In addition, the conductance is enhanced due to asymmetry in the n-p region. We also showed that the Hall plateaus
appear due to equilibration of the quantum Hall edge states in the different regions with electron and hole type charge carriers.

Our approach can be extended to long-range disorder due to charge impurities, and also to the case of spin dependence of the electronic
transport with ferromagnetic-gate in bilayer graphene nanoribbon sheets.

\section{Acknowledgement}

We acknowledge useful discussions with A. G. Moghaddam and H. Hassanian. We would also like to acknowledge M. J{\"a}{\"a}skel{\"a}inen for carefully reading our manuscript. A. Q. has been supported partially by IPM grant.

\newpage
\begin{figure}
\includegraphics[width=7.0cm]{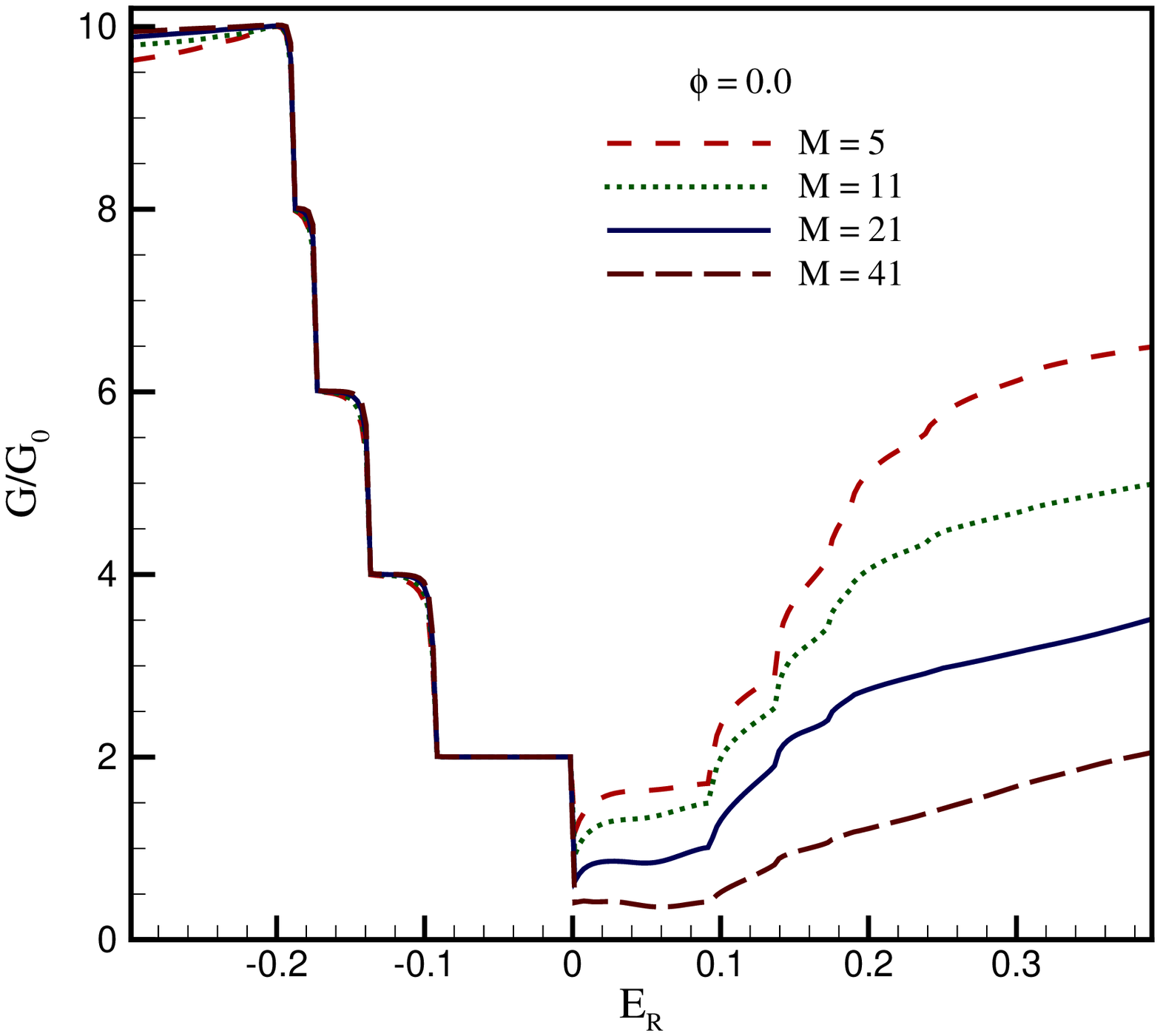}
\includegraphics[width=7.0cm]{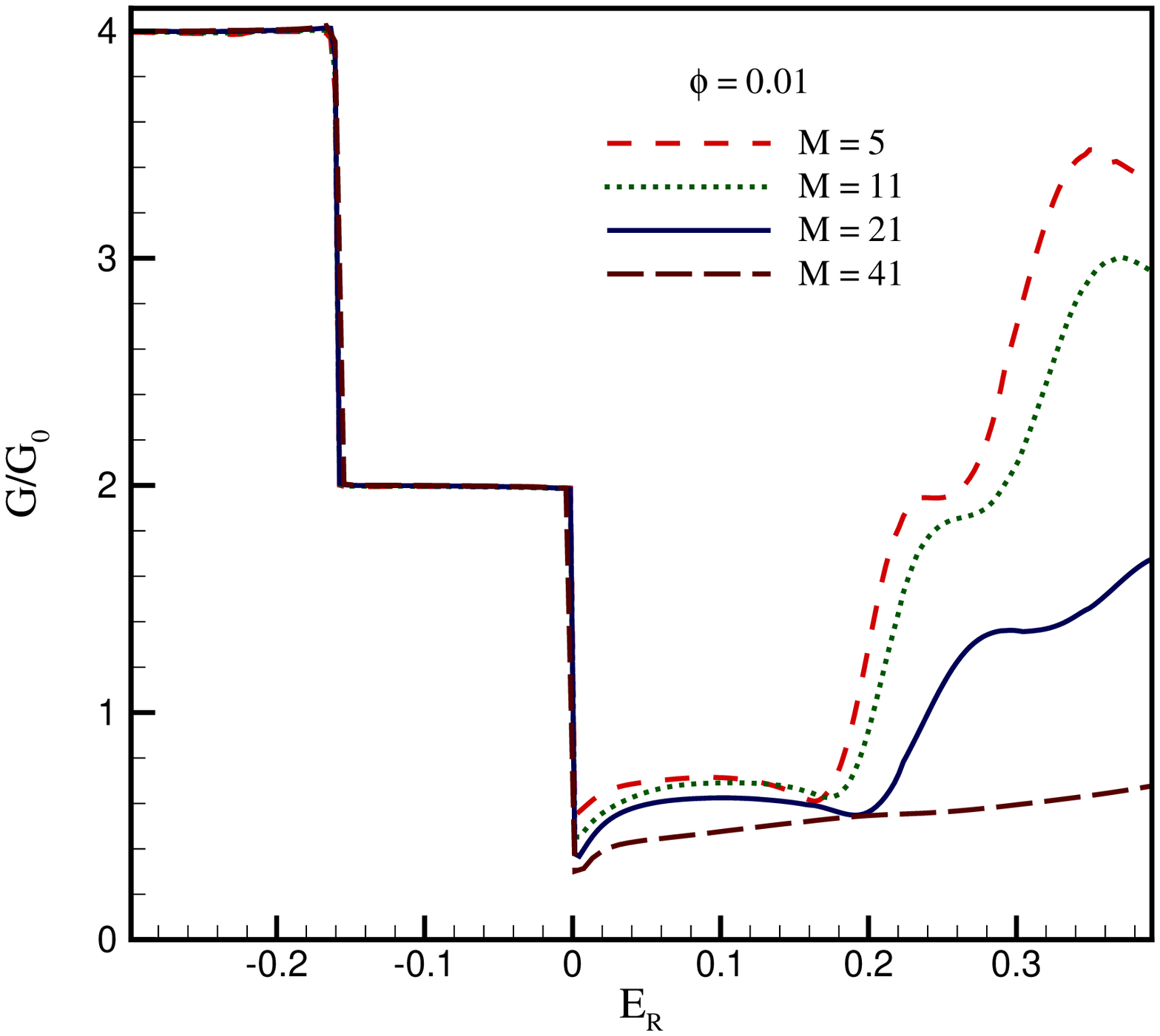}
\caption{(Color online) Conductance of a clean zBGNR as a function
of $E_R$ for various lengths with $E_L=-0.2$ for $\Delta=0.0$ at
$\phi=0$ and $\phi=0.01$.}
\end{figure}

\begin{figure}
\includegraphics[width=7.0cm]{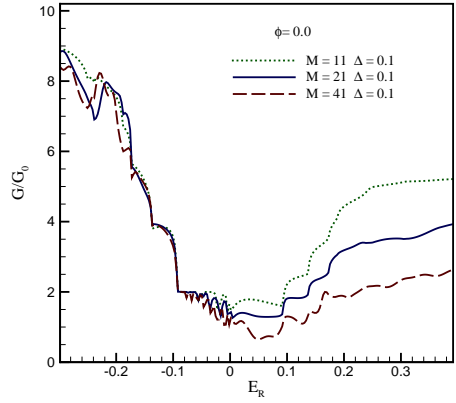}
\caption{(Color online) Conductance of a clean zBGNR with the
asymmetry between two layers, as a function of $E_R$ for various
lengths with $E_L=-0.2$ at zero-magnetic field. }
\end{figure}

\begin{figure}
\includegraphics[width=7.0cm]{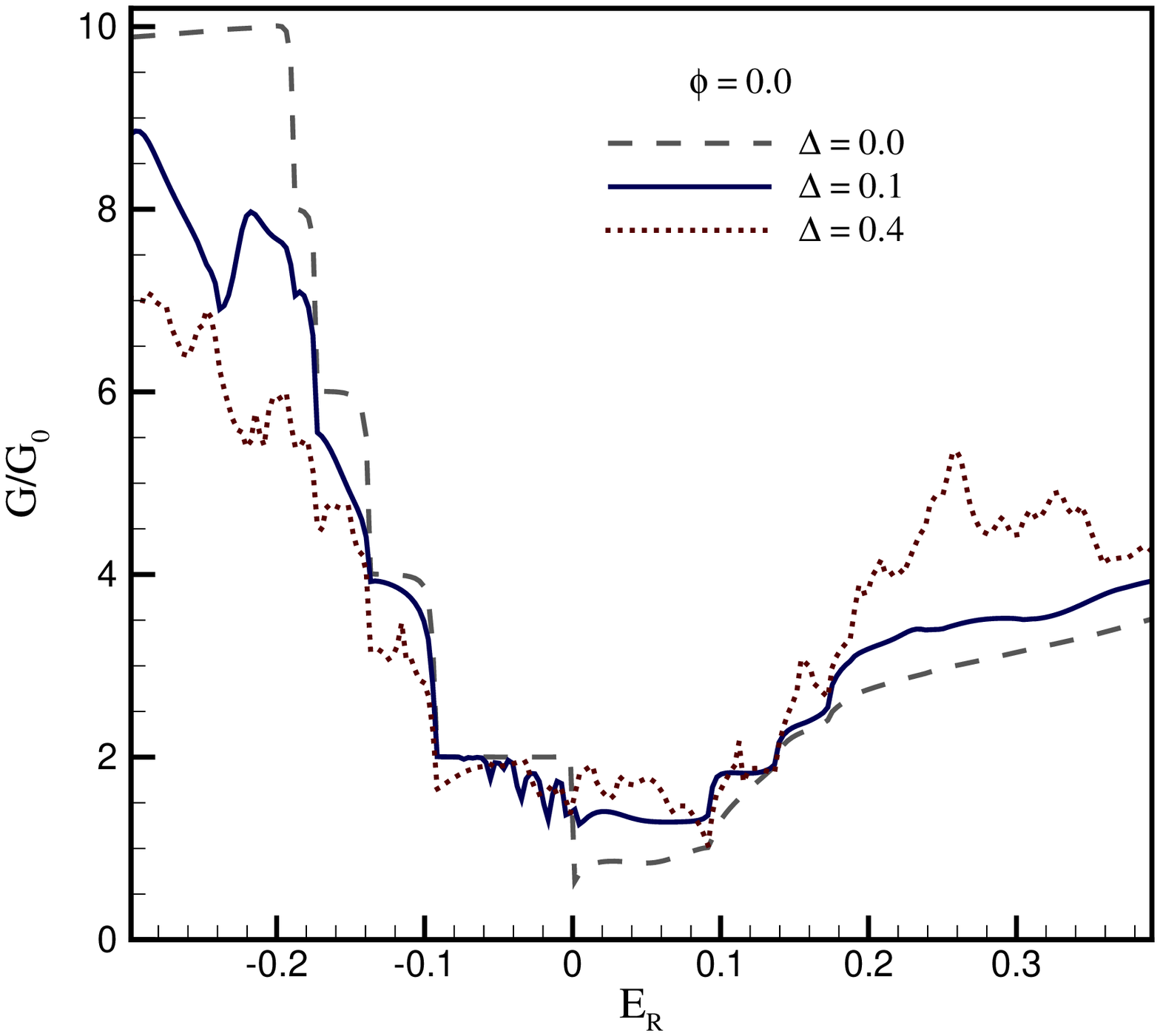}
\includegraphics[width=7.0cm]{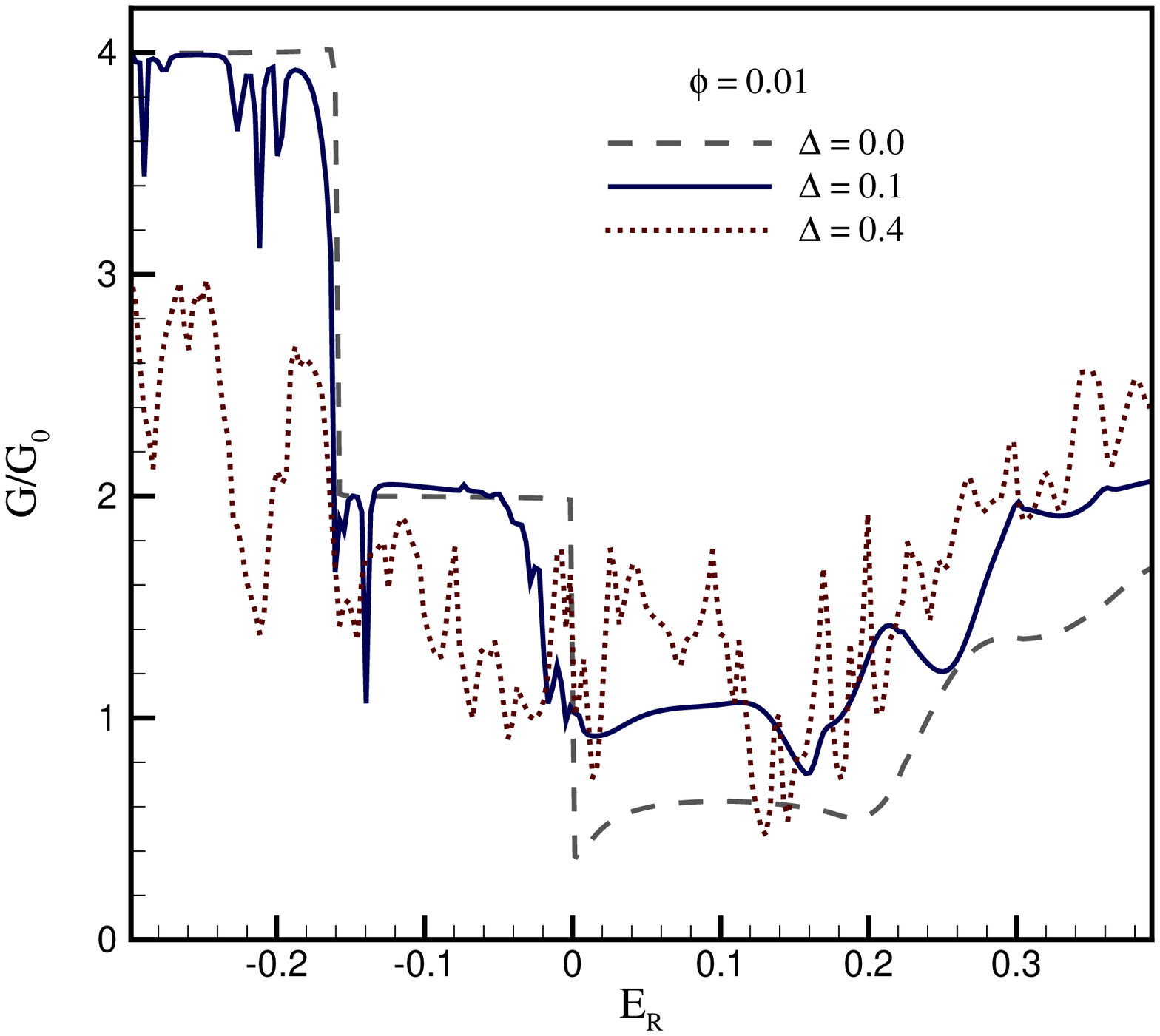}
\caption{(Color online) Conductance of a zBGNR as a function of
$E_R$ for $M=21$, $E_L=-0.2$, and various $\Delta$ values at $\phi=0$ and $\phi=0.01$.}
\end{figure}

\begin{figure}
\includegraphics[width=7.0cm]{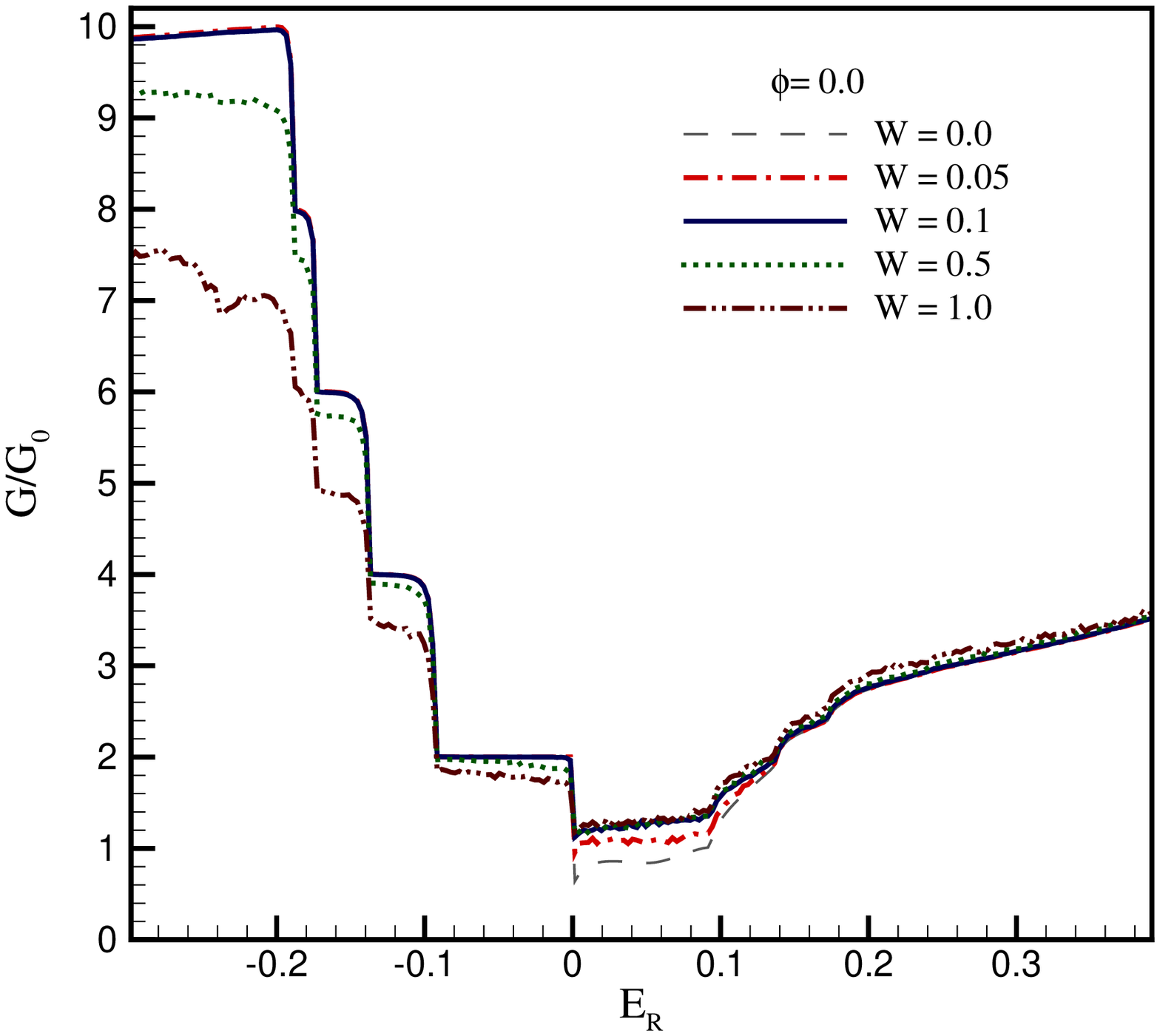}
\includegraphics[width=7.0cm]{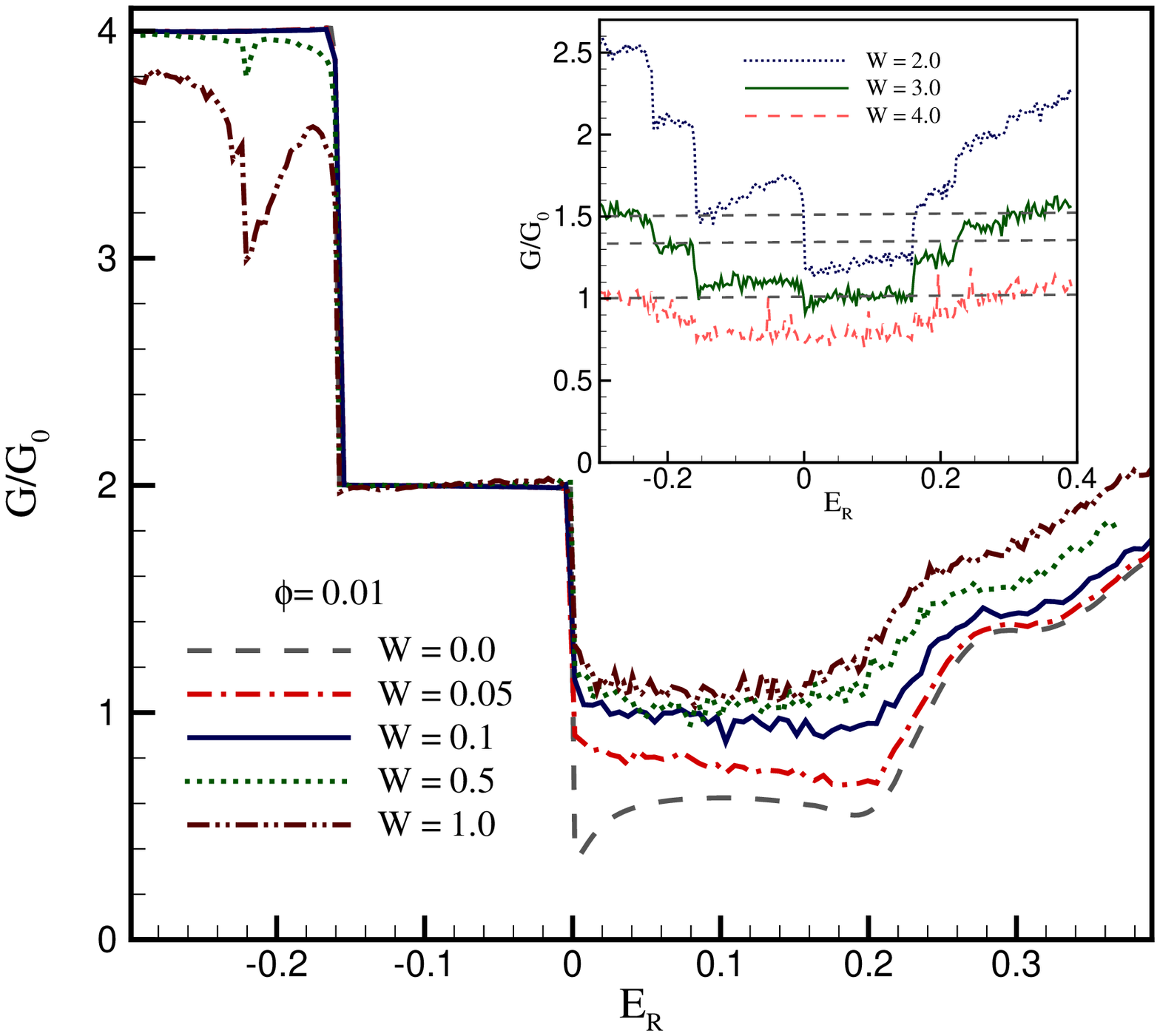}
\caption{(Color online) Conductance of a zBGNR as a function of
$E_R$ for $M=21$, $E_L=-0.2$, and various on-site disorder strengths at
$\phi=0$ and $\phi=0.01$.}
\end{figure}

\begin{figure}
\includegraphics[width=7.0cm]{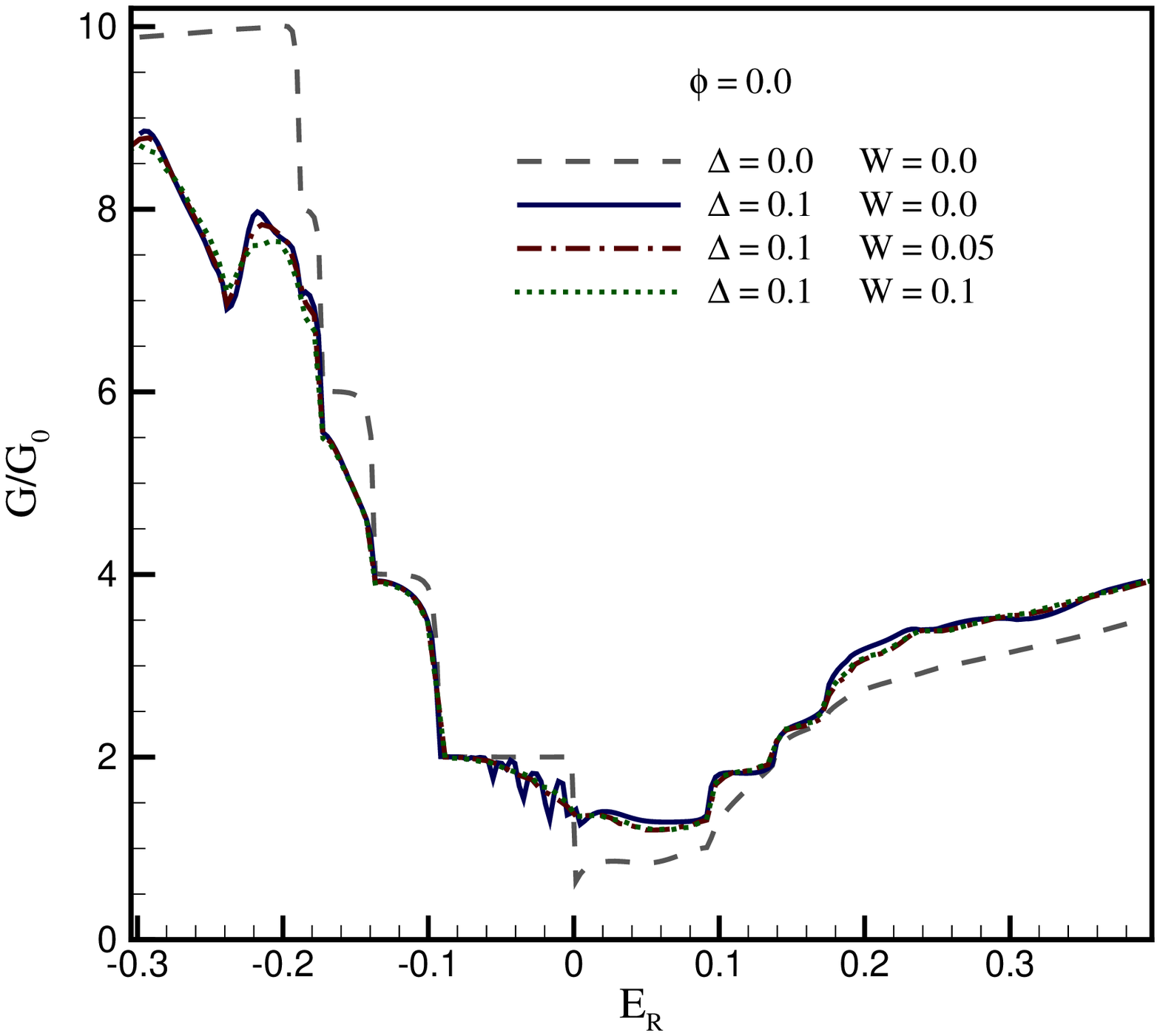}
\includegraphics[width=7.0cm]{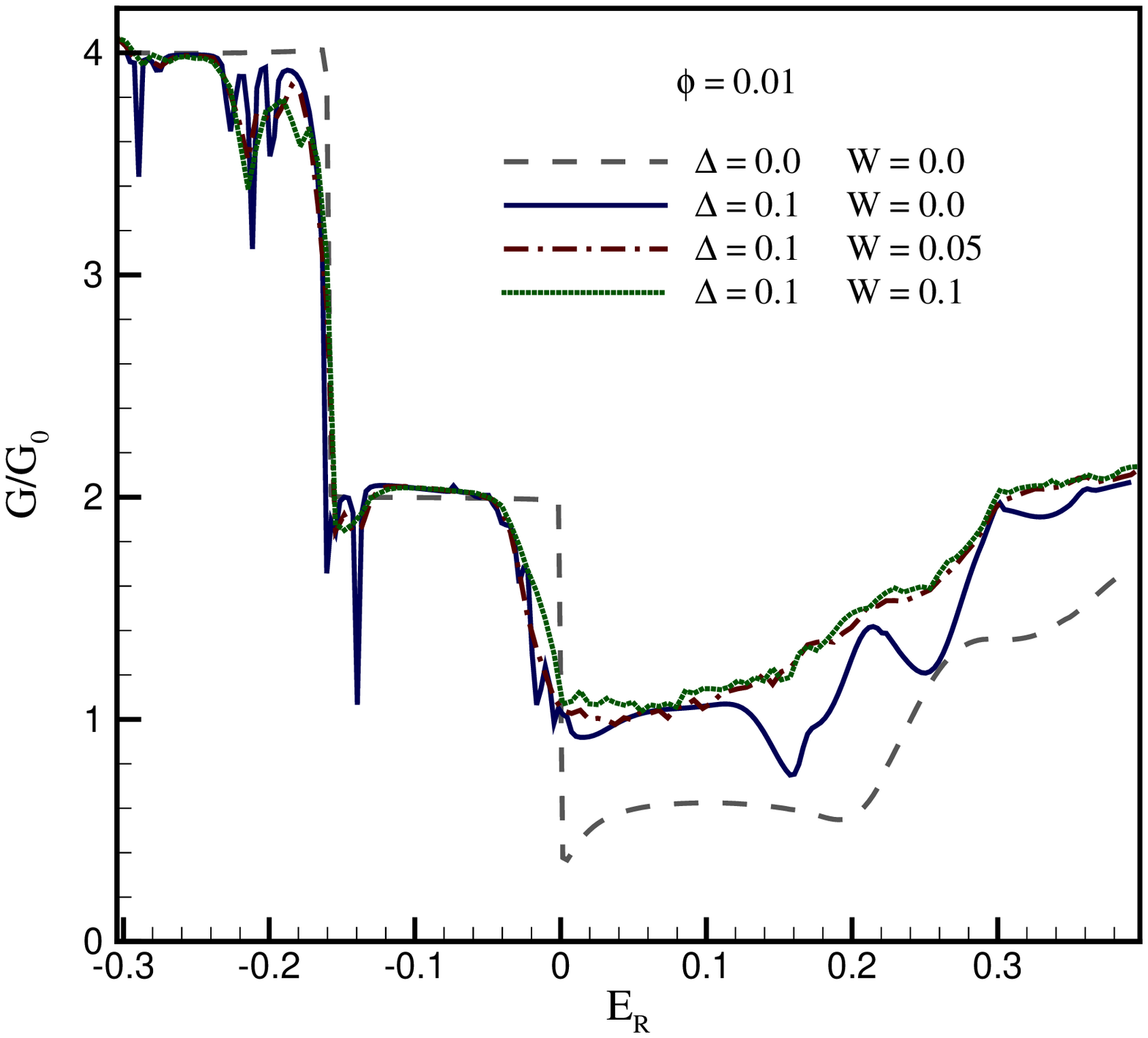}
\caption{(Color online) Conductance of a zBGNR as a function of
$E_R$ for $M=21$, $E_L=-0.2$, $\Delta=0.1$ and various on-site
disorder strengths at $\phi=0$ and $\phi=0.01$
.}
\end{figure}

\begin{figure}
\includegraphics[width=7.0cm]{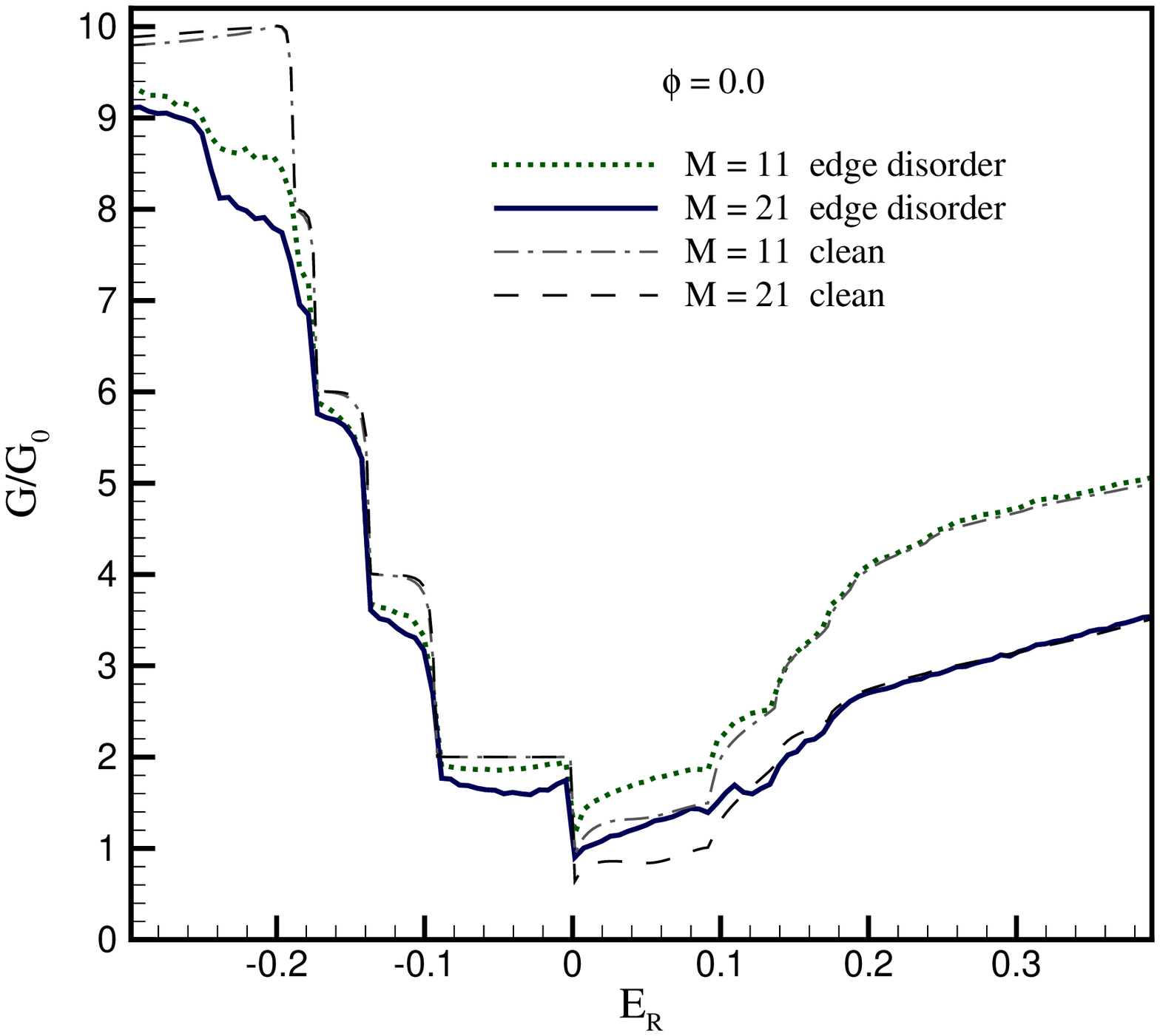}
\includegraphics[width=7.0cm]{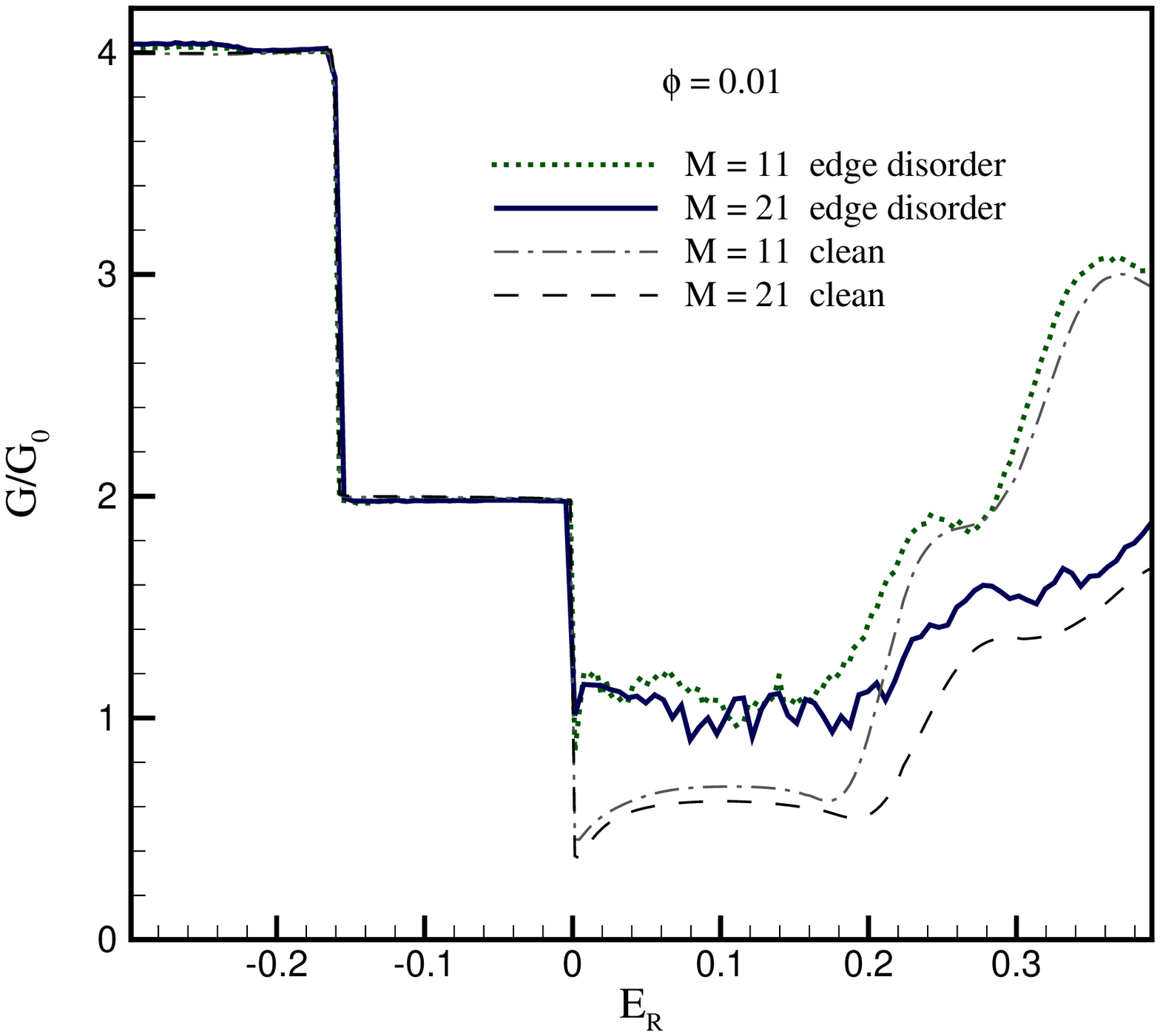}
\caption{(Color online) Edge disordered conductance of a zBGNR as a
function of $E_R$ in comparison with the clean system in the various
values of the lengths for $E_L=-0.2$ at $\phi=0$ and
$\phi=0.01$.}
\end{figure}

\end{document}